\documentclass[12pt]{article}
\usepackage[]{psfig}
\usepackage{latexsym}
\newcommand{\be}{\begin{equation}}
\newcommand{\ee}{\end{equation}}
\newcommand{\ba}{\begin{eqnarray}}
\newcommand{\ea}{\end{eqnarray}}

\begin{document}

\title{Non BPS Noncommutative Vortices}
\author{G.S.~Lozano$^{\,a,b}$\thanks{Associated with CONICET}~ ,
E.F.~Moreno$^{\,c *}$,
M.J.~Rodr\'\i guez$^{\,c}$\thanks{ANPCyT}\\and \\
F.A.Schaposnik$^{\,c}$\thanks{Associated with CICPBA}\\~\\
{$^{a}$\normalsize\it Department of Mathematics, Imperial College London,}\\
{\normalsize\it 180 Queen's Gate, London SW7 2BZ, United Kingdom}\\
{$^{b}$\normalsize\it Departamento de F\'\i sica, FCEyN,
Universidad de Buenos Aires,} \\
{\normalsize \it  Pab.1, Ciudad Universitaria,
 Buenos Aires, Argentina}\\
{$^{c}$\normalsize\it Departamento de F\'\i sica, Universidad Nacional
de La Plata}\\
{\normalsize\it C.C. 67, 1900 La Plata, Argentina}
}
\date{\today}
\maketitle

\begin{abstract}
{ We construct exact vortex solutions to the equations of motion
of the Abelian Higgs model defined in non commutative space,
analyzing in detail the properties of these solutions beyond the
BPS point. We show that our solutions behave as smooth
deformations of vortices in ordinary space time except for parity
symmetry breaking effects induced by the non commutative parameter
$\theta$. }

\end{abstract}

\newpage
\section{Introduction}
The study of noncommutative solitons and instantons -finite energy
or finite action solutions to the classical equations of motion of
noncommutative field theories- has been a field of intense
activity after the revival of interest in these theories   in
connection with strings and brane dynamics (see
\cite{rep1}-\cite{repf} and references therein). In fact, the
first explicit instanton solutions that were constructed in four
dimensional Yang-Mills theory \cite{NS} strongly influenced
developments in string quantization \cite{SW}. Concerning
solitons, not only the noncommutative counterparts of vortex,
monopoles and other localized solutions in ordinary space were
constructed but regular stable solutions which become singular in
the commutative limit were also discovered \cite{GMS1}-\cite{HT}.

Most of these solitons correspond to selfdual/anti-selfdual (BPS)
solutions which are more  simple to obtain than  those arising
from the Euler Lagrange (EL) equations of motion. Moreover, in
even dimensional spaces, calculations can be simplified by
exploiting the connection between noncommutative Moyal product in
configuration space and a Hilbert space representation which
realizes noncommutativity in terms of creation and annihilation
operators acting on a Fock space.

Among the BPS soliton solutions that have been obtained in this
way, particular interest has attracted the construction of
noncommutative BPS vortices - static solutions of the
noncommutative version of the Abelian Higgs model, both when the
gauge field dynamics is governed by Maxwell and/or Chern-Simons
actions \cite{JMW}-\cite{BLP}. The moduli space of these BPS
vortices has been studied in detail \cite{BLP},\cite{T} showing an
interesting phase diagram  with a critical point at some value of
the dimensionless parameter resulting from the combination of the
gauge coupling constant, the scalar expectation value and the
noncommutative parameter.

It is the purpose of the present work to investigate solutions to
the EL equations of motion of the noncommutative Higgs model. That
is, to find, apart the already known BPS and non-BPS
noncommutative vortices, new  non-BPS  solutions which are the
noncommutative counterpart  of the regular vortices originally
introduced by Nielsen and Olesen \cite{NO} and numerically
constructed in \cite{JR}.

The paper is organized as follows. In section 2 we present the
model and establish our conventions. Then, in section 3, we
propose an ansatz to solve the equations of motion in Fock space
reducing the problem to the solution of a system of two coupled
second order recurrence relations. Via the Moyal correspondence
the solutions can be also expressed in ordinary space as an
expansion in Laguerre polynomials with coefficients that can be
computed numerically. We discuss in detail in this section the
properties of vortex solutions with positive magnetic flux and
compare them with those of the commutative case. In section 4 we
present the analogous discussion for negative magnetic flux.
Finally, we summarize our results and conclusions in section 4.

\section{The noncommutative Abelian Higgs model}
We start by defining the Moyal product in four dimensional space-time
in the form
\be
\phi(x) * \chi(x) = \left.
\exp\left( \frac{i\theta^{\mu\nu}}{2}
\partial_\mu^x \partial_\nu^y \right) \phi(x)*\chi(y)
\right\vert_{y=x}
\label{Moyal}
\ee
with $\theta^{\mu\nu}$ a real antisymmetric constant matrix. Since
we are looking for static  solutions we shall take $\theta^{0i} =
0$ ($i,j=1,2,3$) and bring $\theta^{ij}$ into its canonical form
so that
\be
\theta^{12}
=\theta, \;\;\; \theta^{13} = \theta^{23} = 0
\label{thetas}
\ee

Dynamics of the model is governed by the Lagrangian density
\begin{equation}
L=-\frac{1}{4}F_{\mu\nu}*F^{\mu\nu}+
\overline{D_{\mu}\phi}*D^{\mu}\phi-
\frac{\lambda}{4}(\phi\overline{\phi}-\eta^{2})^{2}
\label{lagrangian}
\end{equation}
where
\begin{eqnarray}
F_{\mu\nu}=\partial_{\mu}A_{\nu}-
\partial_{\nu}A_{\mu}-i(A_{\mu}*A_{\nu}-A_{\nu}*A_{\mu})
\label{efe}
\end{eqnarray}
\be
D_{\mu}\phi=\partial_{\mu}\phi-iA_{\mu}*\phi
\ee
Here $A_\mu$ is a $U(1)$ gauge field and $\phi = \phi^1 + i
\phi^2$ a complex scalar. Notice that the gauge coupling constant
has been rescaled to 1 and the covariant derivative has been
chosen as in the ``fundamental'' representation. Other cases
(``antifundamental'' and ``adjoint'' representations) can be
handled in a similar way.

%
Introducing  complex variables
\be
z = \frac{1}{\sqrt 2} (x^1 + i x^2) \; , \;\;\;
\bar z = \frac{1}{\sqrt 2} (x^1 - i x^2)
\ee
the equations of motion read
\begin{eqnarray}
D_{z}D_{\bar z}\phi+D_{\bar z}D_{z}{\phi}&=& \frac{\lambda}{2}(\phi\overline{%
\phi}-\eta^{2})\phi  \nonumber \\
D_{\bar z}F_{{z}{\bar z}}&=&j_{\bar z}  \label{eqs1}
\end{eqnarray}
where
\[
A_z = \frac{1}{\sqrt 2} (A_1 - i A_2) \; , \;\;\;  A_{\bar z} = \frac{1}{%
\sqrt 2} (A_1 + i A_2)
\]
\begin{equation}
j_{\bar z} = -i\left(\vphantom{x^A} ({\partial_{\bar z}}{\phi} )*{\overline{%
\phi}}-{\phi}*({\partial_{\bar z}}{\overline{\phi}}) \right)- ({A_{\bar z}}*{%
\phi})*{\overline{\phi}}-{\phi}* ({\overline{\phi}}*{A_{\bar z}})
\label{cor}
\end{equation}

Noncommutative field theories in two dimensional space can be also handled
by introducing annihilation and creation operators $\hat a$ and ${\hat a}%
^\dagger$ acting on a Fock space,
\begin{equation}
\hat a = \frac{1}{\sqrt \theta} z \; , \;\;\; {\hat a}^\dagger = \frac{1}{%
\sqrt \theta} \bar z  \label{crea}
\end{equation}
\begin{equation}
[\hat a, {\hat a}^\dagger] = 1  \label{alg}
\end{equation}
in terms of which one takes a field $\phi(z,\bar z)$ as an operator $\hat
O_\phi(\hat a, {\hat a}^\dagger)$. The identity
\begin{equation}
\hat O_\phi(\hat a,{\hat a}^\dagger)\hat O_\chi(\hat a,{\hat a}^\dagger)=
\hat O_{\phi*\chi} (\hat a,{\hat a}^\dagger)
\end{equation}
shows that the $*$ product in configuration space becomes the product of
operators in Fock space. Moreover, integration in the plane $(x^1,x^2)$
becomes a trace,
\begin{equation}
\int d^2x F(x^1,x^2) = 2\pi \theta \mathrm{Tr} \hat O_F[\hat a,{\hat a}%
^\dagger]
\end{equation}
With the conventions above, derivatives in the Fock space are given by
\begin{eqnarray}
\partial_{z} = -\frac{1}{\sqrt\theta}[\hat{a}^{\dag},~] \; , \;\;\;
\partial_{\bar z} = \frac{1}{\sqrt\theta}[\hat{a},~]
\end{eqnarray}
so that the EL equations of motion (\ref{eqs1})
become the operator equations
\begin{eqnarray}
&&\!\!\! \!\!\! \frac{1}{\theta}\left(\! [{\hat{a}}^{\dag},[\hat{a},\hat{\phi%
}]] +[\hat{a},[{\hat{a}}^{\dag},\hat{\phi}]]\! \right) -\frac{i}{\sqrt{\theta%
}} [{\hat{a}}^{\dag},\hat{A}_{\bar z}\hat\phi] +\frac{i}{\sqrt{\theta}}\hat{A%
}_{z}[\hat{a},\hat{\phi}] +\left(\!\hat{A}_{z}\hat{A}_{\bar z} +\hat{A}%
_{\bar z}\hat{A}_{z} \!\right)\hat{\phi}  \nonumber \\
&&\!\!\! \!\!\! +\frac{i}{\sqrt{\theta}}[\hat{a},\hat{A}_{z}\hat\phi] -\frac{%
i}{\sqrt{\theta}}\hat{A}_{\bar z}[{\hat{a}}^{\dag}, \hat{\phi}]= - \frac{%
\lambda}{2}(\hat\phi\overline{\hat\phi}-\eta^{2})\hat\phi  \label{ope1}
\end{eqnarray}
\begin{eqnarray}  \label{prerre}
& &\!\!\! \!\!\! \!\!\! \!\!\! \frac{1}{\theta} \left([\hat{a},[{\hat{a}}%
^\dag,\hat{A}_{\bar z}]]+[\hat{a}, [\hat{a},\hat{A}_{z}]]+i\sqrt{\theta} [{%
\hat{a}},[{\hat{A}}_{z},\hat{A}_{\bar z}]+ i\sqrt{\theta} [\hat{A}_{\bar z},[%
\hat{A}_{z},\hat{A}_{\bar z}]] \right)  \nonumber \\
& &\!\!\! \!\!\! \!\!\! \!\!\ \!\!\! + \frac{i}{\sqrt{\theta}} ([\hat{A}%
_{\bar z},[\hat{a}^\dag,\hat{A}_{\bar z}]]+ [\hat{A}_{\bar z},[\hat{a},\hat{A%
}_{z}]] -\frac{i}{\sqrt{\theta}}\left([\hat{a},{\hat{\phi}}] {\overline{\hat{%
\phi}}} -{\hat{\phi}}[\hat{a},{\overline{\hat{\phi}}}]\right)+ %
\hphantom{xxxxxxxxxx}  \nonumber \\
&=& -({A_{\bar z}}{\hat{\phi}}{\overline{\hat{\phi}}}+ {\hat{\phi}}{%
\overline{\hat{\phi}}}{A_{\bar z}})  \nonumber \\
\end{eqnarray}
%

When 
\begin{equation}
\lambda=\lambda_{BPS} = 2  \label{bps}
\end{equation}
-the Bogomol'nyi point- solutions of the  ``BPS'' equations
\begin{eqnarray}
B &=& \eta^2-\hat{\phi} \overline{\hat{\phi}}\; , \;\;\; \,\,\, D_{\bar{z}}
\hat{\phi}=0  \label{sd} \\
-B &=& \eta^2-\hat{\phi} \overline{\hat{\phi}} \; , \;\;\; \,\,\, D_{{z}}
\hat{\phi}=0  \label{asd}
\end{eqnarray}
also solve the Euler-Lagrange equations of motion
(\ref{ope1})-(\ref{prerre}) and saturate a lower bound for the
energy (Eqs. (\ref{sd}) and (\ref {asd}) correspond to the dual
and self-dual cases respectively). Notice that the BPS point
corresponds to the case in which the scalar mass $m_\phi$ and the
vector particle mass $m_A$ ratio, given by
\be \frac{m_\phi^2}{m_A^2} = \frac{\lambda}{2}\; , \ee
is equals to one. In the Ginzburg Landau version of the theory,
the above expression, related to the ratio of the condensate
coherent length and magnetic penetration length signals the
boundary between Type I and Type II superconductors. In the first
case, $\lambda< \lambda_{BPS}$ the range of matter
self-interaction exceeds that of the electromagnetic one leading
to an attractive vortex-vortex interaction while for the second
case the opposite is true.

As mentioned above, exact vortex solutions to the selfdual
eqs.~(\ref{sd}) have been constructed for the whole range $0 \leq
\theta\eta^2 \leq \infty$ \cite{LMS1}. They are the counterpart of
regular vortex solutions to Bogomol'nyi equations in the
commutative case and in fact one can see that they reduce to the
exact solutions found in \cite{dVS} in the $\theta \to 0$ limit.
Concerning the antiselfdual case (\ref{asd}), it has been shown in
\cite{BLP} that solutions exist only in the range $0 \leq
\theta\eta^2 \leq 1$. At the critical point $\theta \eta^2 = 1$,
the BPS solution in this anti-selfdual case coincides with the
fluxon solution discovered in \cite{P}. As for non-BPS solutions
to the Euler-Lagrange equations (\ref{eqs1}), to our knowledge,
the only reported explicit vortex solutions correspond to non-BPS
fluxons \cite{Bak},\cite{BLP}, which exist only in the
anti-selfdual case, are unstable in the range $0 \leq \theta\eta^2
< 1$ and become singular in the commutative limit.

\section{Vortex solutions for positive flux}
Vortex configurations in commutative space take the form \cite{NO}
\begin{equation}
\phi=f(z \overline{z})\frac{z^M}{(z \overline{z})^{\frac{M}{2}}}
\,\,\,\,\,\,\,\,\
A_z=-iM\frac{d(\bar{z} z)}{z}
\label{ansatzcon1}
\end{equation}
\begin{equation}
\phi=f(z \overline{z})\frac{\bar{z}^M}{(z \overline{z})^{\frac{M}{2}}}
\,\,\,\,\,\,\,\,\,\
A_z=iM\frac{d(\bar{z} z)}{z}
\label{ansatzcon}
\end{equation}
for magnetic flux $\Phi$ proportional to  $+M$ and $-M$
respectively. Inspired in (\ref{ansatzcon1})-(\ref{ansatzcon}), we
propose the following ansatz in order to construct exact solutions
to the equations of motion (\ref{ope1})-(\ref{prerre})   for
arbitrary values of the noncommutative parameter $\theta$ and
$\Phi \geq 1$,
\begin{eqnarray}
\hat{\phi} &=&\eta\sum_n f_{n}|n\rangle \langle n+M|\nonumber\\
{\hat A}_z &=&\frac{i}{\sqrt\theta} \sum_n (t_{n}
+\sqrt{n+1})|n+1\rangle \langle n| \label{ansatz}
\end{eqnarray}
(We leave for the next section the case of negative flux).
Plugging the ansatz (\ref{ansatz}) into eqs.(\ref{prerre}) we get the
following recurrence relations for coefficients $f_1$ and $t_1$,
\begin{eqnarray}
2( t_{n} f_{n+1}\; \sqrt{n+1+M} + t_{n-1}f_{n-1}\; \sqrt{n+M}
)+&&\nonumber\\
&&\hspace{-6cm}(t^2_{n}+t^2_{n-1}+2n+2M+1)f_n =-
\frac{{\theta{\eta}^{2}}{\lambda}}{2} f_n({f_{n}}^{2}-1)
\label{unos}
\end{eqnarray}
\begin{eqnarray}
(t_{n+1}^2-2 t^2_{n}+t_{n-1}^2)t_n
={\theta}{\eta}^{2}\left(2f_{n}f_{n+1}\;{\sqrt{n+1+M}}
 +({f_{n}}^{2}
+{f_{n+1}}^{2})t_n \right)
\end{eqnarray}
\begin{eqnarray}
&&f_{1}= -\frac{f_0}{2t_0 \sqrt{1+M}} \left(
 (1+2M) + t_{0}^2 +
\frac{\theta\eta^{2}\lambda}{2}({f_{0}}^{2}-1)\right)\nonumber\\
&& t_1^2=2t_0^2 + \theta \eta^2 \left( (f_1^2+f_0^2) t_0 + 2
\sqrt{1+M} f_0 f_1/t_0  \right) \label{tres}
\end{eqnarray}

Given a value for $f_0$ and $t_0$, one can then determine all
$f_n's$ and $t_n's$ from eqs.(\ref{unos})-(\ref{tres}). The
correct values for $f_0$ and $e_0$ should make
\ba
f_n^2 \to 1  \; , \;\;\;\; t_n \to - {\sqrt{M+n+1}}
\;\;\;\;\;\;\; {\rm as} \;\; n\to \infty\label{as}
\ea
which, as can be seen from ansatz (\ref{ansatz}), correspond to a
scalar field going to its v.e.v. and a gauge field going to a pure
gauge for $r \to \infty$ (the radial variable $r$ is related to
the number operator $\hat N$ in Fock space).

Once all $f_n's$ and $t_n's$ are calculated, one can compute all
relevant quantities. In particular, the vortex  magnetic field can
be computed from
\begin{eqnarray}
&&-i {\hat F}_{z \bar z} \equiv \hat{B}=
\frac{1}{\theta}\sum_{n}{\cal B}_n|n\rangle
\langle n|
\label{B}
\end{eqnarray}
where
\begin{equation}
\hat B = \frac{1}{\theta} \sum_{n} {\cal B}_n |n\rangle \langle n|
\label{bn}
\end{equation}
and
\begin{eqnarray}
{\cal B}_{0}&=&t_0^2-1
\nonumber\\
{\cal B}_n &=& t_n^2 - t_{n-1}^2 - 1 \; , \;\;\;\; n \geq 1
\label{bcoef}
\end{eqnarray}

One can easily calculate (without the need to use the equations of
motion) the magnetic flux $\Phi$
\be
\Phi = 2\pi \theta {\rm Tr} \hat B=2\pi M
\label{flux}
\ee
Expressions in Fock space can be pulled back to configuration
space by using the Moyal mapping. For instance, using the explicit
formula for $|n\rangle\langle n|$ in configuration space in terms
of Laguerre polynomials $L_n$ one ends with
\begin{equation}
B(r) = \frac{2}{\theta} \sum_{n} (-1)^{n} {\cal B}_n
\exp\left(-\frac{r^2}{\theta} \right)  L_n\left(
\frac{2r^2}{\theta}\right) \label{bn1}
\end{equation}

Using the expression for the energy-momentum tensor,
\begin{equation}
T_{00} = 2\pi \theta {\rm Tr} \left( \frac{1}{2} {\hat B}^2 +2
\overline{D_{z}\phi} D_z\phi +2 \overline{D_{\bar z}\phi} D_{\bar
z}\phi+ \frac{\lambda}{4}
\left(\phi\overline{\phi}-\eta^{2}\right)^{2} \right)
\end{equation}
one can write the energy of the vortex configuration in terms of
coefficients $f_n's$ and $t_n's$ as
\begin{eqnarray}
 E^{(M)} &=& 2\pi\sum_{n}\left(
\frac{1}{2\theta}\left(t_n^2 - t_{n-1}^2 - 1\right)^{2} \right. +
\eta^{2}\left( \left(f_{n} t_n + f_{n+1} \sqrt{n+M + 1}
\right)^{2} \right.
\nonumber \\
&&+  \left. \left(f_{n+1} t_n + f_n \sqrt{n+M+1}
 \right)^2
+\frac{\lambda\theta\eta^4}{4}\left( {f_{n}}^2- 1 \right)^2
\right) \label{queseyo}
\end{eqnarray}

For simplicity, we shall first discuss the $M=1$ case and then
comment the case of arbitrary positive integer $M$. Exploring the
whole range of $\theta \eta^2$ and $\lambda$ one finds that vortex
solutions  exist for all the values of $\lambda$ and $\theta
\eta^2$ considered. For $\lambda_{BPS} = 2$, the solution
coincides with that obtained in \cite{LMS1} by solving the  BPS
equations. Concerning the commutative limit (small-$\theta$
regime) we reobtain the exact solution found in \cite{dVS} for
$\lambda = \lambda_{BPS}$ as well as the variational results
obtained in (\cite{JR}) for $1 \leq \lambda \leq 3$. As an
illustration, we show  in figure 1 the $M=1$ vortex magnetic field
as a function of $\theta r$ for $\theta\eta^2 = 2$, and different
values of $\lambda$.  Other ranges of parameters give similar
behavior.

\begin{figure}
\centerline{ \psfig{figure=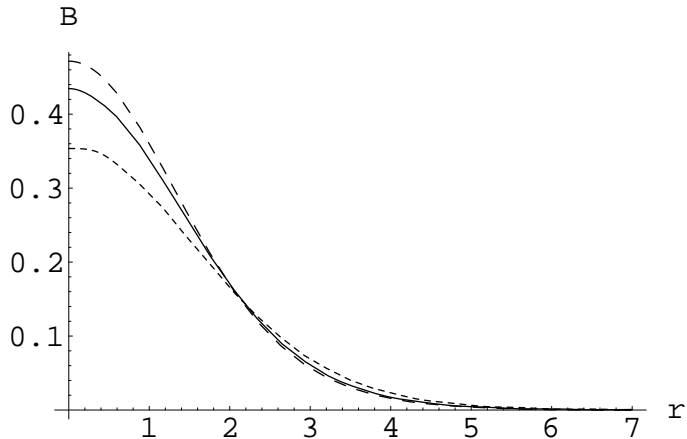,height=6cm,angle=0}}
\smallskip
\caption{\small The $M=1$ vortex magnetic field $B$ as a function
of $r$ (in units of $\eta^2$) for $\theta\eta^2 = 2$ and different
values of $\lambda$: the dotted line corresponds to $\lambda
=0.5$, the solid one to $\lambda = 2$ (the BPS point) and the
dashed one to $\lambda = 8$. \label{fig1} }
\end{figure}

In Fig 2 we show the energy $E^{(1)}$,  as a function of
$\lambda$, for different values of $\theta$. The energy of all
solutions coincide at Bogomol'nyi point ($\lambda =2$) as already
established in \cite{LMS1},
\begin{equation}
\frac{1}{\eta ^{2}}E^{(1)}\left[ \lambda =\lambda _{BPS};\theta \eta ^{2}%
\right] =2\pi
\end{equation}
%

\begin{figure}
\centerline{ \psfig{figure=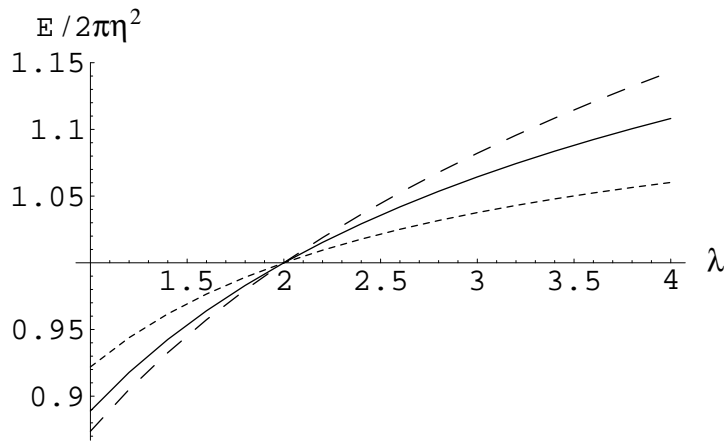,height=6cm,angle=0}}
\smallskip
\caption{\small Energy per unit length  (in units of $\eta^2$) as
a function of $\lambda$: the dashed line corresponds to $\theta
\eta^2 =0.1$, the solid one to  $\theta \eta^2 = 0.5$  and the
dotted one to $\theta \eta^2 = 2$.
\label{fig2} }
\end{figure}

Outside the BPS point the energy is $\theta $ dependent and one
finds, on the one hand
\begin{equation}
\frac{\delta E^{1}(\lambda,\theta)}{\delta \theta} > 0 \;\;\;\;\;\;\;
\lambda<\lambda_{BPS}
\end{equation}
\begin{equation}
\frac{\delta E^{1}(\lambda,\theta)}{\delta \theta}  <0 \;\;\;\;\;\;\;\;
 \lambda>\lambda_{BPS}
\end{equation}
On the other hand, one also has in the whole $\theta $ range
\begin{equation}
\frac{\delta E^{1} (\lambda,\theta)}{\delta \lambda} > 0
\end{equation}

The calculations described above can be easily extended to the
search of vortex solutions with arbitrary positive flux $M$. The
resulting field configurations for $M>1$ are qualitatively similar
to the $M=1$ case.

Nevertheless, it is important in this case, to compare the energy
of the $M$-vortex $E^{(M)}$ with $E^{(1)}$. We show in Fig.~3 the
energy of a $M=2$ vortex compared with twice the energy for an
$M=1$ vortex as a function of $\lambda$ for fixed $\theta$ (
$\theta \eta^2 = 2$). In complete analogy with the commutative
case, there is  a  crossover at the Bogomol'nyi point
$\lambda_{BPS}$ signaling that  for  $\lambda > \lambda_{BPS}$ it
is energetically favorable for a $M=2$-vortex configuration to
decay into two $M=1$ vortices. This behavior indicates that, as in
the commutative case, vortices attract (repel) each other for
values of the coupling constant below (above) the Bogomol'nyi
point. This behavior remains the same for all values of $\theta
\eta^2$ investigated indicating that the character of
attraction/repulsion is unaffected by the value of the parameter
$\theta \eta^2$. Of course, at $\lambda = \lambda_{BPS}$ vortices
do not interact (the stress tensor vanishes \cite{dVS}). In this
case one has $ E^{(2)}[\lambda_{BPS}] = 2 E^{(1)}[\lambda_{BPS}]$.

\begin{figure}
\centerline{ \psfig{figure=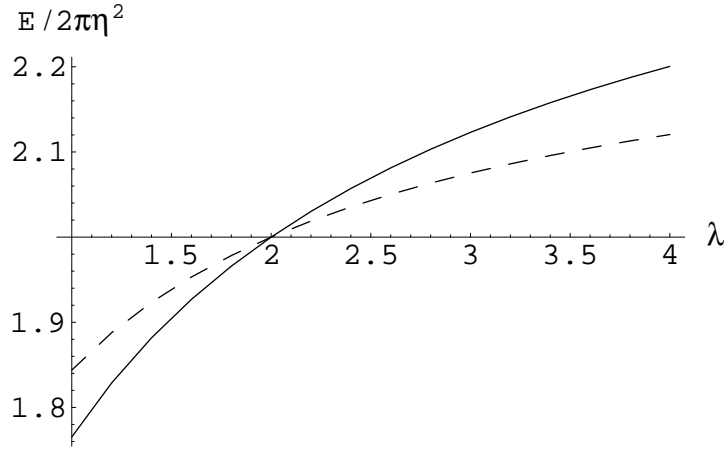,height=6cm,angle=0}}
\smallskip
\caption{\small The energy of an $M=2$ vortex as a function of
$\lambda$ (solid line) compared with that corresponding to twice
the energy of an $M=1$ vortex (dashed line), for $\theta \eta^2
=2$ \label{fig4}}
\end{figure}

Let us end this section with a comment about the accuracy of our
numerical computations. When solving the recursive relations that
define the solutions for  the coefficients $f_n$ and $t_n$, we
have truncated the Fock space to a given value of $n$. Since the
recursive relations are highly nonlinear, it is very difficult to
have a controlled management of the  errors due to that
truncation. However, we can have an estimate of the error in the
computation of the energy by comparing the numerical result at the
Bogol'nyi point $\lambda = \lambda_{BPS}$  (for $M=1,2$, for
example) with the exact analytical results. We found that for the
range of values of $\theta \eta^2$ considered, the error is less
than $10^{-5}$. Our numerical analysis suggests that this estimate
of the error can be extrapolated to the values of $\lambda$
considered in the article.

\section{Vortex solutions for  negative flux}
Since the noncommutativity of space breaks the parity invariance
of the theory, negative flux solutions cannot be obtained from the
positive flux ones by a parity transformation, as in the
commutative case. Negative flux solutions have then to be studied
separately. Thus, instead of ansatz (\ref{ansatz}) one has to
look, in the case of negative magnetic flux, for configurations in
the form
\begin{eqnarray}
{\hat{\phi}}&=&\eta\sum_{n}{f_{n}|n+M\rangle \langle n|}
\label{higgsneg}\\
{\hat{A}}_{z}&=&\frac{i}{\sqrt\theta}\sum_{n} (t_n +\sqrt{n+1})
|n+1\rangle \langle n| \label{ansatzneg}
\end{eqnarray}
\ba f_n^2 \to 1  \; , \;\;\;\; t_n \to -{\sqrt{M+n-1}}
\;\;\;\;\;\;\; {\rm as} \;\; n\to \infty\label{as2} \ea
where $M$ is again a positive integer, $M>0$, this leading to a
negative magnetic flux $\Phi/(2\pi) = -M$.

For simplicity, we present in detail the case $M=1$ but the
analysis goes the same for arbitrary $M$. Using (\ref{ansatzneg}),
the equations of motion (\ref{prerre}) lead to the recurrence
relations for $n>1$

\begin{eqnarray}
&&2\left( t_{n+1} f_{n+1}\; \sqrt{n+1} +  t_n f_{n-1}\;
\sqrt{n}\right) + \left(t^2_{n+1}+ t^2_{n}+2n+1\right)f_n
\nonumber \\
&&=-\frac{{\theta{\eta}^{2}}{\lambda}}{2} f_n({f_{n}}^{2}-1)
\end{eqnarray}
\begin{eqnarray}
(t_{n+1}^2-2 t^2_{n}+t_{n-1}^2) t_n
={\theta}{\eta}^{2}\left(2f_{n}f_{n-1}{\sqrt{n}}
 +({f_{n}}^{2}
+{f_{n-1}}^{2})t_n \right)
\end{eqnarray}
\begin{eqnarray}
&&f_{1}= -\frac{f_0}{2 t_1} \left(1+t_0^{2} +t_1^{2} +
\frac{\theta\eta^{2}\lambda}{2}
({f_{0}}^{2}-1)\right)\nonumber\\
&&t_{1}= \sqrt{2 t_0^2+\theta \eta^2 f_0^2}
\end{eqnarray}

Again, once all $f_n's$ and $t_n's$ are calculated, one can
compute the vortex magnetic field, magnetic flux and energy (since
the ansatz for the gauge field is the same as in the positive flux
case, the magnetic field is again given by
eqs.(\ref{bn}),\ref{bcoef}).

The expression for the energy for a $\Phi/(2\pi) = -1$
configuration takes the form
\begin{eqnarray}
 E^{(M)}&=& 2\pi\sum_{n}\left(
\frac{1}{2\theta}\left(t_n^2 - t_{n-1}^2 - 1\right)^{2} +
\eta^{2}\left( \left(f_{n} t_{n+1} + f_{n+1} \sqrt{n+1}\right)^{2}
\right.\right.
\nonumber \\
&& + \left.
 \left(f_{n} t_n + f_{n-1} \sqrt{n}
 \right)^2 \!
+\frac{\lambda\theta\eta^4}{4}\left( {f_{n-1}}^2- 1 \right)^2
\right) \label{queseyo2}
\end{eqnarray}
(the summation goes from $n=0$ to $n=\infty$ with the proviso that
coefficients with negative subindex vanish).

As shown in \cite{P},\cite{Bak} and \cite{BLP}, there exist in this
case a solution with magnetic flux $\Phi/2\pi = -1$ (a ``fluxon") of
the form,
\begin{eqnarray}
\phi^{fl} &=& \eta \sum_{n=0} |n+1\rangle \langle n| \nonumber\\
A_z^{fl} &=& \frac{i}{\sqrt{2\theta}} \sum_{n=0} \left( \sqrt{n+1}
- \sqrt{n} \right) |n+1\rangle \langle n| \label{flu1}
\end{eqnarray}
Indeed, within this ansatz
\begin{eqnarray}
-B^{fl} &=& \frac{1}{\theta} |0\rangle\langle 0 |  \nonumber\\
D_z \phi^{fl} &=& D_{\bar z} \phi^{fl} = 0 \nonumber\\
 \eta^2 - \phi^{fl}\bar \phi^{fl}   &=& \eta^2
|0\rangle\langle 0 |
\end{eqnarray}
By direct substitution, it is then immediate to show that this
configuration satisfies the EL  equations of motion for all value
of the parameters. The energy of  the fluxon solution (\ref{flu1})
is
\be
\frac{E^{fl}}{2\pi \eta^2} =  \frac{1}{2} \left
 ( \frac{1}{\theta \eta^2} + \frac{\lambda}{2}
\theta \eta^2 \right)
\label{enerflux}
\ee
Nevertheless, a more careful study reveals that this solutions are
locally stable only for $\theta \eta^2>1$. Moreover, they are BPS
saturated only when $\lambda=2$ and $\theta \eta^2=1$ (Note that
for $\theta\eta^2 = 1$ and $\lambda = \lambda_{BPS} = 2$ the
energy does correspond to the BPS bound, $E^{fl} = E_{BPS} = 2\pi
\eta^2$).

Since BPS solutions still can be found for $\lambda=2$ and $\theta
\eta^2<1$ by considering an ansatz of the form (\ref{ansatzneg})
and solving the BPS equation,  the question that  arises concerns
the existence and properties of non BPS solutions for $\theta
\eta^2<1$. In order to answer this question we have investigated
the numerical solutions to the recurrence relations in different
ranges  of $\theta\eta^2$ and $\lambda$.

\begin{figure}
\centerline{\psfig{figure=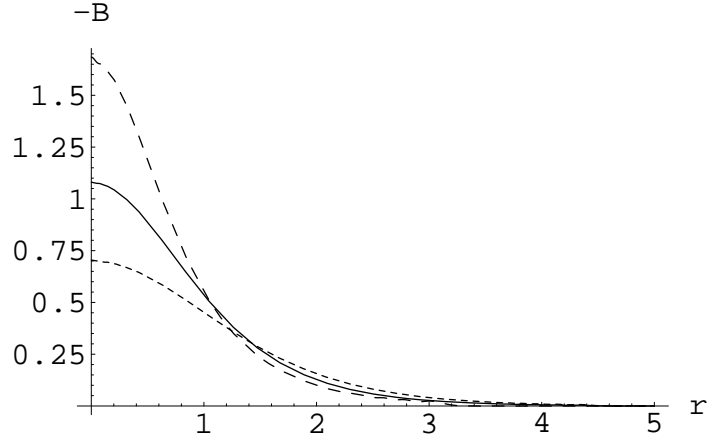,height=6cm,angle=0}}
\smallskip
\caption{\small The magnetic field as a function  of  $r$
 for  $\theta\eta^2 = 0.1$ and different values of  $\lambda$:
 the dotted line corresponds to $\lambda =0.5$,
the solid one to $\lambda = 2$
 and the dashed one to  $\lambda = 8$. \label{fig5} }
\end{figure}

As an illustration, we show in Fig. \ref{fig5} the magnetic field
as a function of $r$ for $\theta \eta^2 = 0.1$ and different
values of $\lambda$. We have calculated numerically the magnetic
flux for this configuration confirming that it corresponds to one
unit of flux.
%

\begin{table}[!hbp]
\begin{small}
\begin{center}
\begin{tabular}{|c|c|c|c|}
\hline \hline
 $\theta\eta^{2}$ & $E[\lambda=0.5]$ & $E[\lambda=2]$ & $E[ \lambda=8]$ \\
\hline
0.1&0.751&1.000&1.400\\
\hline
0.3&0.735&1.000&1.572\\
\hline
0.5&0.716&1.000&1.801\\
\hline
0.8&0.675&1.000&2.206 \\
\hline
0.9&0.654&1.000&2.352\\
\hline \hline
\end{tabular}
\end{center}
\end{small}
\smallskip
\caption{\small The energy of vortex with magnetic flux
$\Phi/2\pi= -1$ for different values of $\theta \eta^2$ and
$\lambda$. \label{table5} }
\end{table}

\begin{table}[!hbp]
\begin{small}
\begin{center}
\begin{tabular}{|p{15mm}|p{25mm}|p{25mm}|p{25mm}|}
\hline \hline
 $\theta\eta^{2}$ & $E_{\Phi=-1}$ & $E_{\Phi=1}$\\
\hline
0.1&0.755&0.762\\
\hline
0.3&0.735&0.775\\
\hline
0.5&0.716& 0.785 \\
\hline \hline
\end{tabular}
\end{center}
\end{small}
\smallskip
\caption{\small Vortex and anti-vortex energies (in units of
$2\pi\eta^{2}$) for $\lambda = 0.5$. \label{table10} }
\end{table}

\begin{figure}
\centerline{ \psfig{figure=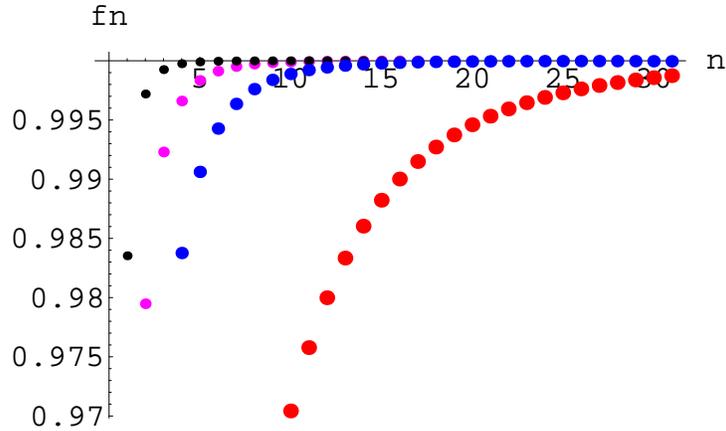,height=6cm,angle=0}}
\smallskip
\caption{\small The coefficients $f_n$ in the Higgs field
development for the $\Phi/2\pi = -1$  solutions  (\ref{higgsneg})
as a function of $n$ for different values of $\theta$. Size of
dots decreases as $\theta\eta^2$ goes from
$\theta\eta^2=0.1$ to
$\theta\eta^2= 0.9$.
\label{fig8}}
\end{figure}

One can compare the values for the energy given in the Table
\ref{table5} with those resulting from formula (\ref{enerflux})
for fluxons to conclude that the energy of the solutions we have
presented in the range $0 \leq \theta\eta^2 <1$ is lower than that
of the (unstable) non-BPS fluxon. Moreover, the energy of our
vortex solution tends to the value of the fluxon solution energy
for $\theta\eta^2 \to 1$. We show in  Fig. \ref{fig8} the behavior
of $f_n$ as $\theta \eta^2\to1$. Indeed at $\theta = 1$ all
$f_n's$ and $t_n's$ for our solutions coincide with those of the
fluxon solutions which,   from that critical value of $\theta$ on
remain as the only non-trivial solutions.

It is interesting also to notice that the asymmetry between vortex
and anti-vortex configurations manifests in the energy splitting
between vortex-antivortex configurations for non zero values of
$\theta \eta^2$ as shown in Table \ref{table10}. Moreover, the
behavior of the energy with $\theta$ is the opposite: for the
vortex the energy increases (decreases) with $\theta$ if
$\lambda<2$ ($\lambda>2$) while for the antivortex the energy
decreases (increases) with $\theta$ if $\lambda<2$ ($\lambda>2$).

\section{Summary and discussion}

In this paper we have examined vortex solutions in the Abelian
Higgs model in non-commutative space, focussing on the properties
of these solutions beyond the BPS point previously considered in
\cite{JMW}, \cite{Bak}, \cite{LMS1}, \cite{BLP}.

Previous to our investigations, the only known non-BPS solutions
were fluxons \cite{P}, \cite{Bak},  negative flux solutions which
are stable only for $\theta \eta^2>1$. These configurations, even
though they are non-BPS in the sense that they do not  satisfy the
duality equations, share some properties with BPS solutions,
namely, their energy saturates a topological bound and is linear
in the flux. Moreover, in the $\theta \to 0$ commutative limite
they correspond to singular configurations (with a
$\delta$-function source).

We have constructed here non-BPS solutions of positive flux with
arbitrary values $\theta \eta^2$  and also negative flux
solutions, in this last case in the range $0\leq \theta \eta^2
<1$. Unlike the fluxon case mentioned above, no simple analytical
expressions of these solutions are available. One has instead
expressions like eq.(\ref{bn1}) so that the properties of the
solutions have to be investigated numerically (as it happens in
the commutative case, both for BPS and non-BPS solutions
\cite{dVS}-\cite{JR}).

The solutions presented here behave in most ways as smooth
deformations of vortices in commutative space. For instance, their
energy is an increasing function of $\lambda$ and is a linear
function of the flux only at the BPS point. Indeed, we have shown
that $E^{(M)}-ME^{(1)}>0$ for $\lambda>\lambda_{BPS}$ suggesting
that in this case, the $M$-vortex configuration should be unstable
towards the formation of a Abrikosov-type vortex lattice in
analogy with Type II superconductors. Notice though that solutions
in non-commutative space differ from solutions in ordinary space
time as a result of parity breaking which manifests itself as a
breaking of symmetry between vortex and anti-vortex
configurations. We have illustrated this fact by comparing the
energies of $E^{(1)}$ and $E^{(-1)}$ as a function of $\theta$.


\vspace{.8 cm}
\noindent\underline{Acknowledgements}: We would like to thank
D.Correa for helpful comments. E.F.M. would like to thanks the
Physics Department of West Virginia University for the hospitality
extended to him while part of this work was done. This work  was
partially supported by UNLP, UBA, CICBA, CONICET, ANPCYT (PICT
grant 03-05179) Argentina and ECOS-Sud Argentina-France
collaboration (grant A01E02). E.F.M. is partially supported by
Fundaci\'on Antorchas, Argentina. GSL is partially supported by
EPSRC grant GR/R70309.


\end{document}